\documentclass[journal=apchd5,manuscript=article]{achemso}
\usepackage[utf8]{inputenc}
\usepackage{graphicx}
\usepackage{amsmath}

\usepackage{xr}
\makeatletter

\newcommand*{\addFileDependency}[1]{
\typeout{(#1)}
\@addtofilelist{#1}
\IfFileExists{#1}{}{\typeout{No file #1.}}
}\makeatother

\newcommand*{\myexternaldocument}[1]{%
\externaldocument[#1:]{#1}%
\addFileDependency{#1.tex}%
\addFileDependency{#1.aux}%
}

\myexternaldocument{SI}

\title{Single-mode emission by phase-delayed coupling between nano-lasers
}
\author{T. V. Raziman}
\affiliation{Blackett Laboratory, Department of Physics, Imperial College London, London SW7 2AZ, UK}
\alsoaffiliation{Department of Mathematics, Imperial College London, London SW7 2AZ, UK}
\author{Anna Fischer}
\affiliation{Blackett Laboratory, Department of Physics, Imperial College London, London SW7 2AZ, UK}
\alsoaffiliation{IBM Research Europe - Z\"{u}rich, R\"uschlikon 8803, Switzerland}
\author{Riccardo Nori}
\affiliation{Blackett Laboratory, Department of Physics, Imperial College London, London SW7 2AZ, UK}
\author{Anthony Chan}
\affiliation{Blackett Laboratory, Department of Physics, Imperial College London, London SW7 2AZ, UK}
\author{Wai~Kit~Ng}
\affiliation{Blackett Laboratory, Department of Physics, Imperial College London, London SW7 2AZ, UK}
\author{Dhruv Saxena}
\affiliation{Blackett Laboratory, Department of Physics, Imperial College London, London SW7 2AZ, UK}
\author{Ortwin Hess}
\affiliation{School of Physics and CRANN Institute, Trinity College Dublin, Dublin 2, Ireland}
\affiliation{Blackett Laboratory, Department of Physics, Imperial College London, London SW7 2AZ, UK}
\author{Korneel Molkens}
\affiliation{Photonics Research Group, Ghent University - Imec, 9052 Gent, Belgium}
\alsoaffiliation{Physics and Chemistry of Nanostructures Group, Department of Chemistry, Ghent University, 9000 Gent, Belgium}
\alsoaffiliation{Center for Nano- and Biophotonics, Ghent University, 9052 Gent, Belgium}
\author{Ivo Tanghe}
\affiliation{Photonics Research Group, Ghent University - Imec, 9052 Gent, Belgium}
\alsoaffiliation{Physics and Chemistry of Nanostructures Group, Department of Chemistry, Ghent University, 9000 Gent, Belgium}
\alsoaffiliation{Center for Nano- and Biophotonics, Ghent University, 9052 Gent, Belgium}
\author{Pieter~Geiregat}
\affiliation{Physics and Chemistry of Nanostructures Group, Department of Chemistry, Ghent University, 9000 Gent, Belgium}
\alsoaffiliation{Center for Nano- and Biophotonics, Ghent University, 9052 Gent, Belgium}
\author{Dries Van Thourhout}
\affiliation{Photonics Research Group, Ghent University - Imec, 9052 Gent, Belgium}
\alsoaffiliation{Center for Nano- and Biophotonics, Ghent University, 9052 Gent, Belgium}
\author{Mauricio Barahona}
\affiliation{Department of Mathematics, Imperial College London, London SW7 2AZ, UK}
\email{m.barahona@imperial.ac.uk}
\author{Riccardo Sapienza}
\affiliation{Blackett Laboratory, Department of Physics, Imperial College London, London SW7 2AZ, UK}
\email{r.sapienza@imperial.ac.uk}

\begin{document}
\maketitle

\begin{abstract}
    Near-field coupling between nanolasers enables collective high-power lasing but leads to complex spectral reshaping and multimode operation, limiting the emission brightness, spatial coherence and temporal stability. Many lasing architectures have been proposed to circumvent this limitation, based on symmetries, topology, or interference.
    We show that a much simpler and robust method exploiting phase-delayed coupling, where light exchanged by the lasers carries a phase, can enable stable single-mode operation.
    Phase-delayed coupling changes the modal amplification: for pump powers close to the anyonic parity-time (PT) symmetric exceptional point, a high phase delay completely separates the mode thresholds, leading to single mode operation.
    This is shown by stability analysis with nonlinear coupled mode theory and stochastic differential equations for two coupled nanolasers and confirmed by realistic semi-analytical treatment of a dimer of lasing nanospheres.
    Finally, we extend the mode control to large arrays of nanolasers, featuring lowered thresholds and higher power.
    Our work promises a novel solution to engineer bright and stable single-mode lasing from nanolaser arrays
    with important applications in photonic chips for communication and lidars.
\end{abstract}

\section{Introduction}

Integrated nanolasers find applications in diverse fields including optical communication~\cite{wang_2017, li_2022}, on-chip computing~\cite{feldmann_2021, skalli_2022, zhou_2023}, and lidars~\cite{zhou_2023, yang_2023, doylend_2012,hulme_2015}.
Many of these applications, however, require substantial output power in a stable single mode.
Whereas subwavelength nanolasers allow stable single-mode operation but are limited in gain, larger high-power nanolasers support multiple spatial and spectral modes, resulting in fluctuations in emission wavelength and power~\cite{fischer_1996, marciante_1998, bittner_2018}. 

Coupling many single-mode nanolasers is not a solution to increase the output power, stability or functionalities, as it leads to complex spectral reshaping and multimode operation which limits the emission brightness, spatial coherence and temporal stability.
Approaches to suppress additional modes in large and collective nanolasers have explored topology, e.g. periodic nanostructures to create photonic crystals and topological lasers~\cite{choi_2021, contractor_2022, yoshida_2023}, symmetry~\cite{hokmabadi_2019}, interference between bright and dark modes, as for bound states in the continuum~\cite{hwang_2021,zhong_2023} and geometrical perturbations~\cite{bogdanov_2015, noh_2019, tiwari_2022}.
However, these methods require high nanofabrication accuracy, hindering practical applications.

The non-hermitian interaction between coupled lasers can be used to achieve single-mode lasing by operating near the exceptional point (EP) where parity-time (PT) symmetry is broken~\cite{liertzer_2012, brandstetter_2014, feng_2014, hodaei_2015, ozdemir_2019, parto_2021}.
Although coalescing of eigenmodes at the EP prevents multimode operation, practical realisation is limited by its extreme sensitivity to the unavoidable inhomogeneities in realistic systems, often exploited for sensing~\cite{wiersig_2020, li_2023}. Achieving EPs with more than two nanolasers is still an open challenge~\cite{demange_2011, hodaei_2017}.
A generalized \textit{anyonic} PT symmetry can be retrieved through complex non-Hermitian coupling between the lasers, resulting in an anyonic EP~\cite{arwas_2022, takata_2022, li_2024}.
However, strong geometric constraints remain since the coupling needs to match the frequency detuning exactly~\cite{arwas_2022}.
When the non-hermitian coupling becomes purely imaginary (\textit{dissipative}), coupled lasers can be made to synchronize where their frequencies are locked to each other~\cite{ding_2019, moreno_2024}.

\begin{figure}[t]
    \centering
    \includegraphics[scale=1]{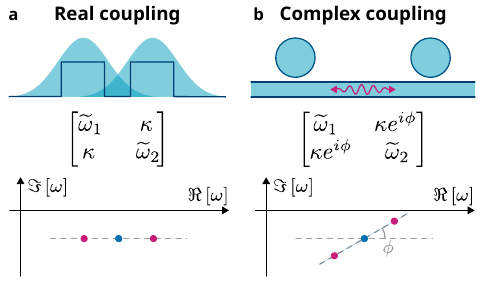}
    \caption{\textbf{Complex coupling}.
    (a) In coupled mode theory (CMT), coupling is typically considered real to account for evanescent interaction between nearby particles.
    This results in the modes of the system splitting in frequency.
    (b) If the particles are far apart and coupled via radiation or waveguide modes, the coupling gains a phase and becomes complex.
    In general, this results in the coupled modes having not only different (real) frequencies but also different losses (imaginary frequencies).
    }
    \label{introduction}
\end{figure}

Here, we use phase-delayed coupling, where light exchanged by the lasers carries a phase~\cite{jiang_2021}, to achieve single-mode lasing in large arrays of coupled nanolasers. 
We employ nonlinear coupled mode theory (CMT) and stability analysis of two coupled nanolasers to demonstrate that, for a high enough phase delay, the modes have different gain and a unique stable lasing mode exists.
We link the transition from multimode to single-mode behaviour with the appearance of anyonic PT symmetry.
We further confirm the single-mode operation with a semi-analytic treatment of two coupled nanospheres.
Finally, we generalise our result to larger arrays, comprising 10 lasers, which also sustain a single mode.
Our results promise a novel direction towards low-complexity fabrication of integrated nanolasers with stable, high-power single-mode operation.

\section{Single-mode lasing from a phase-delayed dimer}

We first show that a system of two frequency-detuned coupled resonators can achieve single-mode lasing by increasing the phase delay of the coupling between them, using linear coupled mode theory (CMT) and a coupling element $\kappa e^{i\phi}$, where $\phi$ is the phase-delay.

Under real coupling ($\phi=0$, Figure~\ref{introduction}a), the modes repel in real frequency [$\textrm{Re}(\omega)$], when the lasers are pumped equally. The threshold pump required for each mode to initiate lasing, [$\textrm{Im}(\omega)$], is very similar to that for the individual lasers, which we can determine from the passive cavity losses. 
Instead, when the coupling is complex ($\phi>0$,  Figure~\ref{introduction}b), a difference in $\textrm{Im}(\omega)$ develops.
Complex coupling arises from phase delay, obtainable through light propagation distances of the order of the wavelength, and can be implemented for example using a waveguide near the individual lasers.

\begin{figure}[t!]
    \centering
    \includegraphics[width=\textwidth]{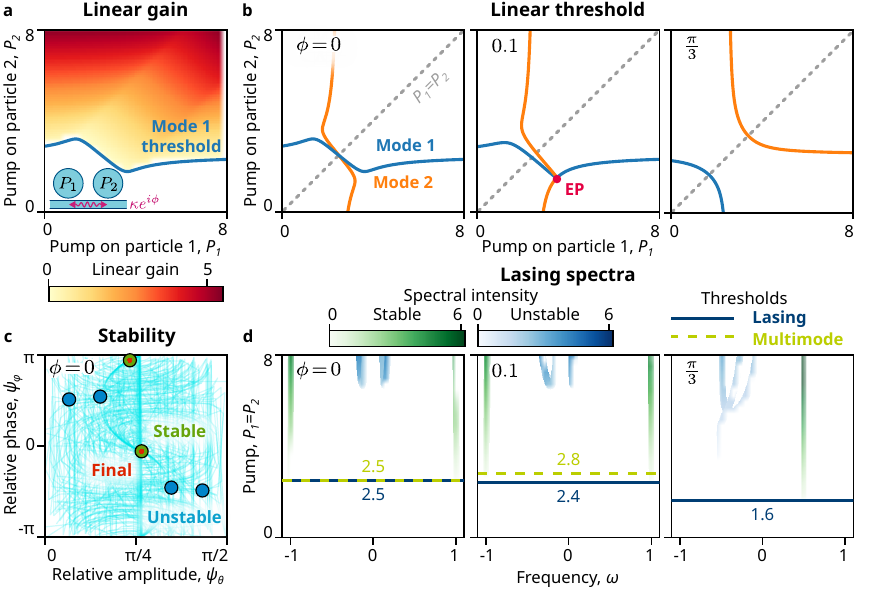}
    \caption{\textbf{Single-mode lasing via complex coupling.}
    (a) Two detuned lasers ($\omega_2 - \omega_1 = 0.2, \gamma_1 = \gamma_2 = 2.5$) interacting through complex coupling ($\kappa e^{i\phi}, \kappa=1, \phi=0$) and pumped with $(P_1,P_2)$.
    Under linear CMT with real coupling ($\phi=0$), described by the 2x2 matrix in Fig.~\ref{introduction}a, two coupled modes arise, which lase for pump values beyond the threshold curves (blue and orange), where their linear gain is positive. Linear gain of one of the modes is shown.
    (b) At $\phi=0$, the threshold curves cross symmetrically. On increasing $\phi$, they become asymmetric and separate from each other at the exceptional point (EP, $\phi=0.1$), yielding a large single-mode region at large delay ($\phi=\pi/3$).
    (c) On evolving the system with random noise, the state trajectories (cyan) only evolve to the stable modes (green) identified by nonlinear CMT.
    (d) Under equal pump ($P_1=P_2$), at $\phi=0$, two modes reach threshold simultaneously.
    On increasing $\phi$, the two modes separate in threshold, with the second mode requiring an even higher pump to reach stability.
    For high values of $\phi$, we obtain \textbf{single-mode lasing operation} where only one mode is stable.
    }
    \label{singlemode}
\end{figure}

When the two coupled lasers are pumped unequally, the threshold of each mode depends on both excitations~\cite{liertzer_2012, brandstetter_2014, fischer_2024}. 
This dependence can be visualised as a threshold curve in the $(P_1, P_2)$ plane as in Figure~\ref{singlemode}a, for $\phi = 0$. The blue curve indicates the threshold of one of the modes (here mode 1), which is now a wavy line (and would have been a straight line for uncoupled lasers). The linear gain increases with the pump $P_i$, as indicated by the colourmap.


In the presence of complex coupling ($\phi>0$, Figure~\ref{singlemode}b), the threshold curves become asymmetric, specifically shifting their intersection away from the $P_1=P_2$ line for increasing $\phi$.
Under equal pump (gray dotted line), one mode reaches its threshold just before the other, enabling a limited range of single-mode operation.
For a specific value of the coupling phase, here $\phi=0.1$, the eigenvectors of the two modes coalesce at an exceptional point (EP) with anyonic PT symmetry.
For $\phi>0.1$, the thresholds separate from each other with one curve becoming convex and the other, concave.
Beyond this, one mode consistently reaches its threshold before the other, leading to a large range of single-mode operation. 

However, above threshold, linear CMT provides an incomplete picture with unphysical exponential growth of mode amplitudes with time.
In real lasers, mode amplitudes are constrained by gain saturation, which we incorporate in nonlinear CMT~\cite{benzaouia_2022, fischer_2024} (Supplementary Section~\ref{SI:sec:method}).
Nonlinearity leads to the emergence of more coupled modes.

We employ Jacobian stability analysis to assess the stability of these modes to identify the modes observable in experiments (Supplementary Section~\ref{SI:sec:stability}).
Under real coupling, only two modes exhibit stability, with equal intensities as we expect from symmetry (Fig.~\ref{singlemode}d).
We confirm the stability of modes using time-domain simulations of the underlying coupled differential equations, starting from zero amplitude and adding random noise (see Supporting Section~\ref{SI:sec:timedomain}).
We consistently observe that the system state (cyan) converges to one of the stable modes (Figure~\ref{singlemode}c, green circles) and never to any unstable mode (blue circles), confirming that the modes assigned stable are the experimentally observable ones.

Nonlinear CMT predicts an extended range of single-mode operation enabled by complex coupling compared to linear CMT.
As the coupling phase $\phi$ increases, the lasing threshold decreases, and the second mode requires a higher pump intensity to attain stability than the linear onset (Figure~\ref{singlemode}d).
At high values of $\phi$, the second mode is never stable, resulting in a single stable lasing mode at all powers above threshold.
We attribute this observation to the significant separation between the threshold curves.
Even with gain saturation, any mode arising from the higher threshold curve will retain enough gain for a mode from the lower curve to emerge and dominate it.
This single-mode operation is robust to pump inhomogenities -- the system has a single stable lasing mode even when the two lasers are pumped unequally (Figure~\ref{SI:modecount}).
Further, single-mode lasing is attained at values of the coupling phase $\phi$ much lower than the studies of synchronisation under dissipative coupling ($\phi=\pi/2$)~\cite{ding_2019, moreno_2024}.

Introducing a high phase delay can thus effectively suppress multimode behaviour in laser dimers, allowing for the sustained operation of a single stable lasing mode.

\begin{figure}[t!]
    \centering
    \includegraphics[scale=1]{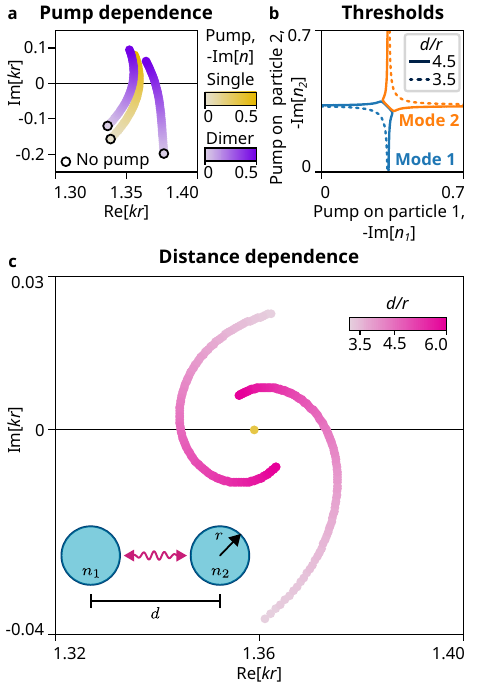}
    \caption{\textbf{Complex coupling in a sphere dimer}.
    (a) When two spheres of radius $r$ separated by $d$ are coupled, the mode of the sphere (brown) splits into two modes (purple) that differ both in frequency and gain.
    On pumping the spheres, one of the coupled modes reaches threshold before the other.
    (b) Threshold curves of the dimer vary in separation as the distance is varied, making one mode have more gain than the other.
    (c) Increasing the distance between the spheres changes both the amplitude and the phase of coupling, resulting in the coupled modes moving spirally around the single sphere mode.
    }
    \label{spheredimer}
\end{figure}

\section{Phase-delayed coupling in realistic systems}

Realistic nanolasers are usually cylindrical~\cite{kwon_2010, yu_2010}, hexagonal~\cite{wang_2013, fischer_2024}, or spherical~\cite{noginov_2009}, while more complex 3d architectures are also starting to be investigated. We validate our model beyond the idealised CMT by investigating the coupling between the lowest-order vector spherical harmonic modes in a dimer of identical spheres based on Mie theory~\cite{Bohren_1998_C4, Chew_1995_AD} (See Supporting section \ref{SI:sec:dimer}).

Phase-delayed coupling can be achieved in coupled sphere nanolasers by adding a physical distance between them. $\phi$ increases with the separation between the spheres, but at the expense of reducing the magnitude of the coupling strength $\kappa$ as the fraction of the scattered light reaching the other sphere reduces.

The simulation in Figure~\ref{spheredimer}a confirms that when two unpumped spheres are coupled, the coupled modes (purple line) have distinct values of real frequency and gain (imaginary part), indicating a complex effective coupling constant.
As the pump is increased in both spheres equally, the gains of the modes increase, until eventually one mode reaches lasing threshold on intersecting the real axis before the other, confirming our observations from CMT.

This effect holds for all pump powers on the two spheres as shown in Figure~\ref{spheredimer}b, where the threshold curve ($\textrm{Im}[kr]=0$) is plotted 
for two intra-sphere distances  $d/r=4.5$ (solid line) and $d/r=3.5$ (dotted line), where $d$ is the ratios between the distance and sphere radius.
For $d/r=4.5$ the threshold curves of the two modes intersect, resembling the prediction from CMT under real coupling, however, for $d/r=3.5$, a noticeable gap emerges between the two threshold curves.
This observation indicates that one mode requires significantly less pump power to reach the lasing threshold compared to the other, aligning with the prediction from CMT under highly complex coupling.

To demonstrate the tunability of the phase, we maintain a fixed and equal pump (corresponding to the threshold of a single sphere) and vary the distance $d$. 
The coupled modes of the system exhibit a spiral trajectory around the frequency of the mode of the single sphere (Figure~\ref{spheredimer}c) due to two key factors.
First, the continuous variation in the complex phase of the coupling makes the coupled modes encircle the single mode and separates them in gain.
Second, the magnitude of the coupling decreases as the spheres move apart, bringing the coupled mode frequencies closer to the single mode frequency.
These results illustrate that manipulating the distance between the spheres in the dimer can effectively tune the phase of the coupling, in turn enabling us to optimize the lasing threshold and achieve single-mode operation.

\section{Single-mode lasing in large arrays of nano-lasers}

\begin{figure}[t!]
    \centering
    \includegraphics[scale=1]{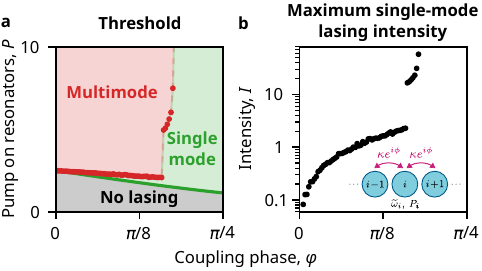}
    \caption{\textbf{Single-mode lasing from large arrays}.
    On increasing coupling phase $\phi$ in a chain of $N=10$ resonators with nearest-neighbour interactions, the threshold of the fundamental mode reduces, and the range and intensity of single-mode operation increase.
    }
    \label{largearray}
\end{figure}

Single-mode operation due to phase-delayed coupling can be generalised to larger arrays of lasers. Here we calculate up to ten coupled lasers in a linear array using stochastic differential equations, with nearest neighbour interactions (Figure~\ref{largearray}). Phase-delayed coupling reduces the lasing threshold (green line), with an effect that increases when more and more lasers are coupled (Figure~\ref{SI:2vs10}).

The pump range permitting single-mode operation in the array also increases with the phase of coupling.
This is due to a combination of the same two factors seen previously in the dimer system: Increasing $\phi$ not only increases the linear gain difference between different modes but also makes the low-gain modes unstable and thus unattainable at low power.
As some modes lose stability entirely at high values of complex phase, multimode onset increases drastically, allowing much higher power of single-mode lasing emission from the system.
Unlike the dimer, the stochastic evolution of the 10-resonator system shows non-eigen multifrequency solutions, but these are also suppressed with complex coupling.



These findings illustrate that phase-delayed coupling can achieve single-mode operation across a broad range of pump powers in large arrays. 
Although multiple modes and complex dynamical solutions exist in such arrays, phase-delayed coupling effectively suppresses them and allows a single mode with the lowest threshold to dominate the system.


\section{Conclusion and outlook}

In conclusion, we have demonstrated the remarkable potential of phase-delayed coupling in achieving tunable single-mode lasing in nanolasers.
Increasing the phase delay between coupled resonators suppresses both multimode eigensolutions and complex dynamical solutions in large arrays, sustaining a single stable lasing mode.
The transition from multimode to single-mode lasing on increasing the phase delay is closely connected to the origin of anyonic PT-symmetry, which is the critical point at which the coupled modes separate in threshold space.
Nonlinear CMT and stability analysis are crucial in describing coupled lasers, showing a more extended range of single-mode operation than predicted by linear CMT.
Our demonstration that phase-delayed coupling makes single-mode lasing possible in realistic systems and large arrays makes this a promising direction for the development of stable and high-power single-mode laser systems.
Extending our analysis to incorporate physical effects such as resonator geometries, realistic semiconductor gain and the phase of coupling through waveguides will provide a more comprehensive understanding of realistic single-mode lasers.
This knowledge will aid in designing lasers for photonic chips for applications such as optical communication, quantum information processing, and LIDARs.

\section{Acknowledgments}
We acknowledge computational resources and support provided by the Imperial College Research Computing Service (http://doi.org/10.14469/hpc/2232).
TVR, DS, and RS acknowledge support from The Engineering and Physical Sciences Research Council (EPSRC), grant number EP/T027258.
AF acknowledges support from the EU ITN EID project CORAL (GA no. 859841).
 O.H. acknowledges financial support from the Science Foundation Ireland (SFI) via grant number 18/RP/6236.
WKN acknowledges the research support funded by the President’s PhD Scholarships from Imperial College London.

\bibliography{references}

\end{document}